\documentstyle[tighten,preprint,amsfonts,%
epsf,eqsecnum,aps,prd]{revtex}
\begin{document}
\draft

\hyphenation{
mani-fold
mani-folds
geo-metry
geo-met-ric
}



\def\BbbR{{\Bbb R}}
\def\BbbZ{{\Bbb Z}}
\def\BbbC{{\Bbb C}}

\def\RPthree{{{\Bbb RP}^3}}
\def\RPtwo{{{\Bbb RP}^2}}

\def\s3{{{S}^3}}

\def\rpdes{{{\Bbb RP}^3{\rm dS}}}
\def\des{{{S}^3{\rm dS}}}

\def\half{{\frac{1}{2}}}
\def\casehalf{{\case{1}{2}}}

\def\evacdes{{|0_{\rm E}\rangle}}
\def\evacrpdes{{|0_{{\Bbb RP}^3{\rm E}}\rangle}}

\def\boulvacdes{{|0_{\rm BdS}\rangle}}

\def\desquadrant{{Q_0}}
\def\rpquadrant{{Q_+}}

\def\statictime{{\sigma}}
\def\estatictime{{\tilde\statictime}}

\def\fieldmass{{\mu}}
\def\fieldeffmass{{\tilde\mu}}

\def\hypergtwoone{{}_{2\!} F_1}

\def\Othree{{\rm O}(3)}
\def\SOthree{{\rm SO}(3)}

\def\Ofourone{{\rm O}(4,1)}
\def\SOfourone{{\rm SO}(4,1)}
\def\Ofour{{\rm O}(4)}
\def\SOfour{{\rm SO}(4)}

\def\arcsinh{\mathop{\rm arcsinh}\nolimits}
\def\arccosh{\mathop{\rm arccosh}\nolimits}
\def\arctanh{\mathop{\rm arctanh}\nolimits}


\preprint{gr-qc/9812056}

\title{The exponential law:
Monopole detectors, 
Bogoliubov transformations, 
and the thermal nature of the 
Euclidean vacuum in $\RPthree$ de~Sitter spacetime}
\author{Jorma Louko\footnote{%
Electronic address:
louko@aei-potsdam.mpg.de. 
Address after September~1, 1999: 
School of Mathematical Sciences, 
University of Nottingham, 
Nottingham NG7 2RD, U.K\null. 
}}
\address{
Max-Planck-Institut f\"ur Gravitations\-physik,
Am~M\"uhlenberg~5, 
D-14476 Golm,
Germany
}
\author{Kristin Schleich\footnote{Electronic address:
schleich@noether.physics.ubc.ca}}
\address{
Department of Physics and Astronomy, 
University of British Columbia, 
\\
Vancouver, BC~V6T~1Z1, Canada}
\date{Revised version, April 1999. 
Published in 
{\it Class.\ Quantum Grav.\ \bf 16} (1999) 2005--2021. 
}
\maketitle
\begin{abstract}%
We consider scalar field theory on the 
$\RPthree$ de~Sitter spacetime ($\rpdes$), which is locally isometric
to de~Sitter space (dS) but has spatial topology $\RPthree$. 
We compare the Euclidean vacua on $\rpdes$ and 
dS in terms of three quantities that are relevant for an 
inertial observer: 
(i)~the stress-energy tensor; 
(ii)~the response of an inertial monopole particle detector; 
(iii)~the expansion of the Euclidean vacuum in terms of many-particle
states associated with static coordinates centered at an
inertial world line. In all these quantities, the 
differences between $\rpdes$ and dS 
turn out to fall off 
exponentially at early and late proper times along the inertial
trajectory. 
In particular, (ii) and (iii) yield 
at early and late proper times in 
$\rpdes$ the usual thermal
result in the de~Sitter Hawking temperature. 
This conforms to what one might
call an exponential law: in expanding locally 
de~Sitter spacetimes, differences due to global topology should fall
off exponentially in the proper time. 
\end{abstract}

\narrowtext

\section{Introduction}
\label{sec:intro}

Observable consequences of the large scale topology of the universe
are a subject of increasing interest. It has been recognized for some
time that a topologically nontrivial universe can produce multiple
images of individually identifiable objects in the sky, and our not
having seen such images sets bounds on the potential scale of
nontrivial topology in our universe 
\cite{soko-shvarts,paal,gott,ellis-schreiber}. 
More recently, it was
recognized that nontrivial topology can also leave an imprint on
the cosmic microwave background, through the quantum
mechanical origin of density inhomogeneities and the subsequent
Sachs-Wolfe effect, and the observational bounds obtained in this way
could in fact be more stringent 
\cite{sss,levin-spots}. 

The purpose of this paper is to explore another situation in which
quantum fields in a curved spacetime feel the large scale topology: We
consider the experiences of an inertial observer coupled to a quantum
field in a spacetime that is locally de~Sitter but has spatial
topology $\RPthree$ instead of the usual~$S^3$. We consider a free
scalar field, and we assume the field to be in the globally regular
vacuum state that is induced by the Euclidean vacuum on de~Sitter
space. As the unconventional spatial topology lies beyond the
cosmological horizon of the inertial observer, this problem
illustrates how quantum fields can probe large scale topology that is
classically unobservable by virtue of topological
censorship \cite{topocen}. Also, as the cosmological horizon of the
observer does not coincide with a bifurcate Killing horizon, this
problem sheds light on the role of the bifurcate Killing horizon in
the thermal effects experienced by an inertial observer in de~Sitter
space \cite{wald-qft}.

We shall find, from the analysis of a monopole particle detector, as
well as from a Bogoliubov transformation between the Euclidean vacuum
and the vacuum natural for the inertial observer, that the experiences
of the inertial observer are not identical to those of an inertial
observer in the Euclidean vacuum in de~Sitter space.  However, in the
limit of early or late proper times along the observer trajectory, the
differences vanish exponentially, and the
experiences of the observer become asymptotically thermal in the usual 
de~Sitter Hawking temperature. 
We also
compute the renormalized stress-energy tensor, finding that it
reduces to that in de~Sitter space in the limit of early and late
times on each inertial trajectory.  These results conform to what
one might call an exponential law: in expanding locally de~Sitter
spacetimes, differences due to global topology should fall off
exponentially in the proper time. 
{}From the viewpoint of the absence of a bifurcate Killing horizon,
qualitatively similar results have 
been previously found on the 
single-exterior eternal black hole known as the
$\RPthree$ geon \cite{louko-marolf-geon} and on the conformal boundary 
of the $(2+1)$-dimensional single-exterior black hole known
as the $\RPtwo$ geon \cite{louko-marolf-btz}. 

The rest of the paper is as follows. In section \ref{sec:spacetimes}
we briefly review the properties of four-dimensional de~Sitter
spacetime, which we denote by~dS, and the quotient construction of a
spacetime, denoted by $\rpdes$, which has the same local geometry but
whose spatial topology is~$\RPthree$. Scalar field theory and the
Euclidean vacua on these spacetimes are introduced in
section~\ref{sec:quantization}, and the stress-energy tensor on 
$\rpdes$ is
evaluated by point-splitting methods. Section 
\ref{sec:bogo-transf} constructs the Bogoliubov transformation on
$\rpdes$, and the particle detector is analyzed in
section~\ref{sec:detector}. Section 
\ref{sec:discussion} contains a brief summary and discussion. 
An evaluation of the
stress-energy tensor by conformal methods, in the special case of a
conformal field, is given in the Appendix. 

We work in Planck units, $\hbar = c = G = 1$. A~metric with signature
$({-}{+}{+}{+})$ is called Lorentzian and a metric with signature
$({+}{+}{+}{+})$ Riemannian. All scalar fields are global sections of
a real line bundle over the spacetime ({\em i.e.,} we do not consider
twisted fields). Complex conjugation is denoted by an overline.

\section{de~Sitter spacetime and $\rpdes$}
\label{sec:spacetimes}

In this section we briefly review the geometry of four-dimensional
de~Sitter spacetime 
(dS) and its quotient space~$\rpdes$. The main purpose
of the section is to establish the notation and to introduce the
coordinate systems that will be used with the quantum field
theory.

\subsection{de~Sitter spacetime}

Four-dimensional de~Sitter space is a 
Lorentzian spacetime form of positive sectional
curvature. 
It can be represented as the hyperboloid
\begin{equation}
H^{-2} = - U^2 + V^2 + X^2 + Y^2 + Z^2
\label{des-hyperboloid} 
\end{equation}
in five-dimensional Minkowski space with the global coordinates 
$(U,V,X,Y,Z)$ and the metric 
\begin{equation}
ds^2 = - dU^2 + dV^2 + dX^2 + dY^2 + dZ^2
\ \ . 
\label{five-mink-metric}
\end{equation}
The parameter $H>0$ is the inverse of 
the radius of curvature of the embedded
hypersurface. 
The spacetime is Lorentzian, and it solves
Einstein's equations with the cosmological constant 
$\Lambda = 3H^2$.
The Ricci scalar is $R = 12 H^2$. The spacetime is globally
hyperbolic, with spatial topology~$S^3$, 
and a global 3+1 foliation is
provided for example by the spacelike hypersurfaces of constant~$U$. 
The (connected component of the)
isometry group is (the connected component of) $\Ofourone$.  We
denote this spacetime by~dS\null.

If $x$ and $y$ denote 
points in~dS, we define 
\begin{equation}
{\cal Z}(x,y) := H^2 \eta_{ab} X^a(x) X^b (y) 
\ \ ,
\label{calZed-def}
\end{equation}
where $X^a(x)$ and $X^a(y)$ are the five-dimensional Minkowski
coordinates of the points on the hyperboloid~(\ref{des-hyperboloid}), 
and $\eta_{ab}$ is the five-dimensional Minkowski
metric~(\ref{five-mink-metric}).
${\cal Z}(x,y)$ is clearly invariant
under the isometries of~dS\null, and it encodes almost all the
isometry-invariant information about the relative location of $x$
and~$y$. In particular, $x$ is on the light cone of $y$ if and only if
${\cal Z}(x,y)=1$. For more
detail, see for example \cite{allen}. 

dS~admits several coordinatizations that are adapted to different
isometry subgroups.  Of relevance to this paper are two: 
hyperspherically symmetric coordinates
and static coordinates. We now exhibit these. 

We introduce on dS the chart $(t,\chi,\theta,\varphi)$ by 
\begin{mathletters}
\begin{eqnarray}
&&U = 
H^{-1} 
\sinh(Ht)
\ \ ,
\\
&&V = 
H^{-1} 
\cosh(Ht) \cos\chi
\ \ ,
\\
&&Z = 
H^{-1} 
\cosh(Ht) \sin\chi \cos\theta
\ \ ,
\\
&&X = 
H^{-1} 
\cosh(Ht) \sin\chi \sin\theta \cos\varphi
\ \ ,
\\
&&Y = 
H^{-1} 
\cosh(Ht) \sin\chi \sin\theta \sin\varphi
\ \ . 
\end{eqnarray}
\end{mathletters}
The metric reads 
\begin{equation}
ds^2 = - dt^2 + H^{-2}\cosh^2\!(Ht) \, d\Omega_3^2
\ \ ,
\label{des-glob-metric}
\end{equation}
where $d\Omega_3^2$ is the metric on the unit three-sphere, 
\begin{equation}
d\Omega_3^2 := 
d\chi^2 + \sin^2\!\chi 
\left( 
d\theta^2 + \sin^2\!\theta \, d\varphi^2 
\right)
\ \ .
\label{threespheremetric}
\end{equation}
The angles $(\chi,\theta,\varphi)$ form a standard set of
hyperspherical coordinates on~$S^3$, and the coordinate singularities
of this chart on $S^3$ can be handled in the standard way. When
$(\chi,\theta,\varphi)$ is understood in this extended sense as as a
global coordinatization of~$S^3$, the chart $(t,\chi,\theta,\varphi)$
and the metric (\ref{des-glob-metric}) are global on dS with
$-\infty<t<\infty$.
The world lines at
constant $(\chi,\theta,\varphi)$ are timelike geodesics, and
the proper time along them is~$t$. 

The coordinates $(t,\chi,\theta,\varphi)$ make manifest the $\Ofour$
isometry subgroup whose orbits are at constant~$t$. Conversely, the
3+1 foliation of dS given by these coordinates is uniquely specified
by the choice of a particular $\Ofour$ isometry subgroup. 

\begin{figure}
\begin{center}
\vglue 3 cm
\leavevmode
\epsfysize=8cm
\epsfbox{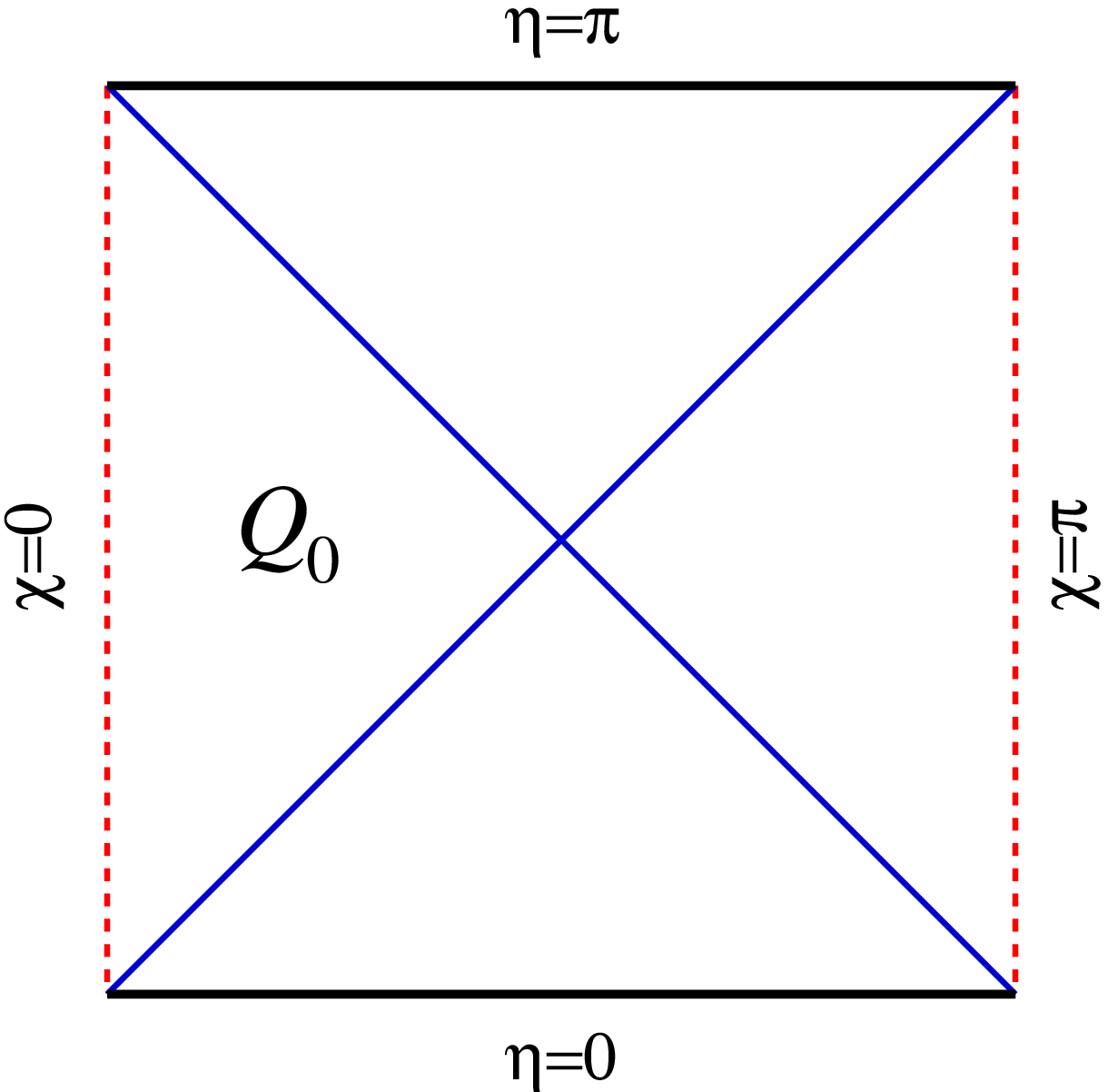}
\end{center}
\vskip 3 cm
\caption{%
A~conformal diagram of~dS\null. The coordinates shown are
$(\eta,\chi)$, and those suppressed are $(\theta,\varphi)$. For
$0<\chi<\pi$, each point in the diagram 
represents a suppressed two-sphere of
radius $H^{-1}\sin\chi/(\sin\eta)$; at $\chi=0$ and $\chi=\pi$, 
each point in the diagram
represents a point in the spacetime. 
The quadrant~$\desquadrant$, 
covered by the static coordinates 
$(\statictime,r,\theta,\varphi)$, is at 
$\cos\chi > |\cos\eta|$. 
The involution~$J$, introduced in the text, consists
of the reflection $(\eta,\chi) \mapsto (\eta,\pi-\chi)$
about the vertical axis, followed by the antipodal map
$(\theta,\varphi) \mapsto (\pi-\theta,\varphi+\pi)$ 
on the suppressed two-sphere.}
\label{fig:des}
\end{figure}

It is useful to introduce the conformal time~$\eta$, 
\begin{equation}
\eta := 2 \arctan (e^{Ht}) 
\ \ ,
\end{equation}
which takes the values $0 < \eta < \pi$.  
As $\cosh(Ht) = 1/\sin\eta$, the
metric (\ref{des-glob-metric}) takes the form
\begin{equation}
ds^2 = 
\frac{1}{H^2 \sin^2\!\eta}
\left[ 
- d\eta^2 + d\Omega_3^2
\right]
\ \ .
\label{des-conf-metric}
\end{equation}
The coordinates $(\eta,\chi)$ are therefore appropriate for a 
conformal diagram in which $(\theta,\varphi)$ are suppressed. Such a 
conformal diagram is shown in Figure~\ref{fig:des}.

We now turn to the static coordinates. 
Let $\desquadrant$ be the quadrant 
$V>|U|$ of~dS. In $\desquadrant$, we introduce the chart 
$(\statictime,r,\theta,\varphi)$ by 
\begin{mathletters}
\label{static-chart}
\begin{eqnarray}
&&U = 
H^{-1} 
\sqrt{1 - H^2r^2}
\sinh(H\statictime)
\ \ ,
\label{static-chart-U}
\\
&&V = 
H^{-1} 
\sqrt{1 - H^2r^2}
\cosh(H\statictime)
\ \ ,
\label{static-chart-V}
\\
&&Z = 
r \cos\theta
\ \ ,
\\
&&X = 
r \sin\theta \cos\varphi
\ \ ,
\\
&&Y = 
r \sin\theta \sin\varphi
\ \ .
\end{eqnarray}
\end{mathletters}
The metric takes the static form 
\begin{equation}
ds^2 = 
- \left( 1 - H^2 r^2 \right) d\statictime^2 + 
{\left( 1 - H^2 r^2 \right)}^{-1} dr^2 + r^2 d\Omega_2^2
\ \ ,
\label{static-metric}
\end{equation}
where $d\Omega_2^2$ is the metric on the unit two-sphere, 
\begin{equation}
d\Omega_2^2 
:= 
d\theta^2 + \sin^2\!\theta \, d\varphi^2
\ \ .
\end{equation}
For $0<r<H^{-1}$, the set $(r,\theta,\varphi)$ forms a standard set of
three-dimensional polar coordinates, and the coordinate singularity at
$r=0$ and at the singularities of the spherical coordinates
$(\theta,\varphi)$ on the two-spheres of constant $r$ can be handled
in the standard way.  When $(r,\theta,\varphi)$ is understood in this
extended sense as as a global coordinatization of~$\BbbR^3$, with
$0\le r<H^{-1}$, the metric (\ref{des-glob-metric}) with
$-\infty<\statictime<\infty$ is global on~$\desquadrant$. 
In the chart
$(\eta,\chi,\theta,\varphi)$, $\desquadrant$ is the region $\cos\chi >
|\cos\eta|$, as shown in the conformal diagram in
Figure~\ref{fig:des}. The transformation between the 
charts reads
\begin{mathletters}
\label{transf-charts}
\begin{eqnarray}
&&Hr = 
\frac{\sin\chi}{\sin\eta}
\ \ ,
\\
&&H\statictime =
- \arctanh \left(\frac{\cos\eta}{\cos\chi}\right)
\ \ .
\end{eqnarray}
\end{mathletters}

$\desquadrant$ has topology~$\BbbR^4$. As seen from
Figure~\ref{fig:des}, it is globally hyperbolic, and the
hypersurfaces of constant $\statictime$ are Cauchy surfaces for
$\desquadrant$ (but not for~dS). 
The curve at $r=0$ is a
timelike geodesic in~dS, 
and $\statictime$ is the proper time along this
geodesic: the static coordinates 
$(\statictime,r,\theta,\varphi)$ are centered around the world line of 
an inertial observer at $r=0$. The boundary of $\desquadrant$, at
$r\to H^{-1}$, is the cosmological 
horizon for this observer, 
and the Killing vector~$\partial_\sigma$, 
which is timelike in~$\desquadrant$, 
becomes null at the horizon. 
The horizon has
therefore an infinite redshift. 

In the quadrant $V> -|U|$, or $\cos\chi < - |\cos\eta|$, a
similar static chart can be introduced with the obvious
modifications. The future and
past quadrants, $U > |V|$ and $U < - |V|$, can be
covered by charts in which (\ref{static-chart-U}) and 
(\ref{static-chart-V}) are replaced by 
\begin{mathletters}
\begin{eqnarray}
&&U = 
\pm
H^{-1} 
\sqrt{H^2r^2 - 1}
\cosh(H\statictime)
\ \ ,
\label{qstatic-chart-U}
\\
&&V = 
H^{-1} 
\sqrt{H^2r^2 - 1}
\sinh(H\statictime)
\ \ 
\label{qstatic-chart-V}
\end{eqnarray}
\end{mathletters}
with the upper (lower) sign in (\ref{qstatic-chart-U}) in the future
(past) quadrant.  The metric in the future and past quadrants takes
the form (\ref{static-metric}) with $r>H^{-1}$.

As any timelike geodesic in dS can be mapped to any other by an
isometry, a static metric of the form (\ref{static-metric}) can be
introduced in a quadrant of the spacetime centered around any timelike
geodesic. The horizon of the static coordinates is in this sense
observer-dependent.

\subsection{The quotient spacetime $\rpdes$}
\label{subsec:rpdes}

On the five-dimensional Minkowski space~(\ref{five-mink-metric}),
consider the map 
\begin{equation}
{\tilde J}: (U, V, X, Y, Z) \mapsto (U, -V, -X, -Y, -Z)
\ \ . 
\label{Jtilde-def}
\end{equation}
We denote by $J$ the map that ${\tilde J}$ induces on~dS\null. In the
coordinates $(\eta, \chi, \theta, \varphi)$, we have
\begin{equation}
J: (\eta, \chi, \theta, \varphi) 
\mapsto 
(\eta, \pi-\chi, \pi-\theta, \varphi+\pi)
\ \ .
\label{J-def}
\end{equation}
$J$~is an involutive isometry, it acts without fixed points and 
properly discontinuously, and it preserves both space and time
orientation. The quotient space ${\rm dS}/J$ is a space and time
orientable Lorentzian manifold. 
We refer to this quotient space as $\RPthree$ de~Sitter space and
denote it by $\rpdes$. 

\begin{figure}
\begin{center}
\vglue 3 cm
\leavevmode
\epsfysize=8cm
\epsfbox{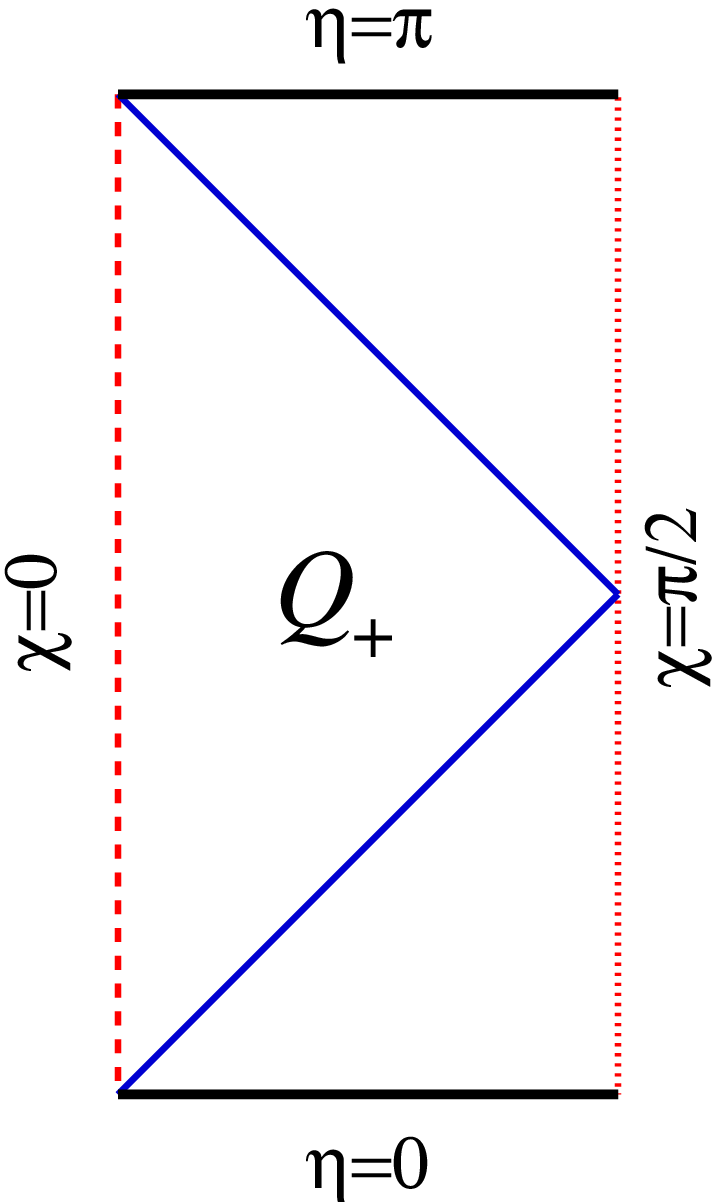}
\end{center}
\vskip 3 cm
\caption{%
A~conformal diagram of~$\rpdes$. 
The region $\chi<\casehalf\pi$ is identical to
that in the diagram of
Figure~\ref{fig:des},
each point with $0<\chi<\casehalf\pi$
representing a suppressed two-sphere in the spacetime, 
and each point
at $\chi=0$ representing a point in the spacetime. 
At $\chi=\casehalf\pi$,
each point in the diagram represents a suppressed~$\RPtwo$.
The region~$\rpquadrant$, 
covered by the static coordinates 
$(\statictime,r,\theta,\varphi)$, is at 
$\cos\chi > |\cos\eta|$.}
\label{fig:rpdes}
\end{figure}

$\rpdes$ is globally hyperbolic, with spatial topology~$\RPthree$. The
chart $(\eta, \chi, \theta, \varphi)$ can be reinterpreted as a global
chart $\rpdes$, provided the angles are understood in the sense of
hyperspherical coordinates on $\RPthree$ and not on~$S^3$: with this
reinterpretation, equation (\ref{des-glob-metric}) gives the global
metric on $\rpdes$. A~conformal diagram in which the coordinates
$(\theta,\varphi)$ are suppressed is shown in Figure~\ref{fig:rpdes}.
As seen in the figure, one can represent $\rpdes$ by taking the region
$0\le\chi\le\casehalf\pi$ of~dS and identifying at $\chi=\casehalf\pi$
the antipodal points on the two-spheres coordinatized by
$(\theta,\varphi)$.

The isometry group of $\rpdes$ is $\BbbZ_2 \times \Ofour$, as induced
by the largest subgroup of $\Ofourone$ 
that commutes with~${\tilde J}$. 
In the coordinates $(\eta, \chi, \theta, \varphi)$ on~$\rpdes$,
the $\Ofour$ factor acts trivially on~$\eta$, while the nontrivial
element of the $\BbbZ_2$ factor acts trivially on the angles and sends
$\eta$ to $\pi-\eta$.  The connected component of the isometry group
is~$\SOfour$. It follows that the $(3+1)$ foliation of $\rpdes$
provided by the coordinates $(\eta, \chi, \theta, \varphi)$ is a
geometrically distinguished one: it is the only foliation in which the
spacelike hypersurfaces are orbits of the connected component of the
isometry group.

As $J$ maps the quadrants $V>|U|$ and $V<-|U|$ of dS onto each other,
these two quadrants project onto a region of $\rpdes$ 
that is isometric to a
single quadrant. We denote this region of $\rpdes$ by~$\rpquadrant$,
and we introduce on it the chart $(\statictime,r,\theta,\varphi)$
induced by the chart (\ref{static-chart}) on~$\desquadrant$. 
Equation (\ref{static-metric}) gives then a
globally-defined metric on~$\rpquadrant$, 
and the chart $(\statictime,r,\theta,\varphi)$ gives an explicit 
isometry between $\rpquadrant$ and~$\desquadrant$. 
The line $r=0$ in
$\rpquadrant$ is a timelike geodesic that is orthogonal to the
distinguished foliation of $\rpdes$. 

{}From the isometries of $\rpdes$ it is immediate that a static metric
of the form (\ref{static-metric}) could be introduced in a
wedge in $\rpdes$ centered around any timelike geodesic orthogonal to
the distinguished foliation. 
It is straightforward to show that a similar static
metric could be introduced also around the timelike geodesics that are
not orthogonal to the distinguished foliation.

\section{Scalar field quantization and the Euclidean vacuum}
\label{sec:quantization}

We now turn to the quantum theory of a real scalar field~$\phi$. In
this section we recall the definition and some characteristic
properties of the Euclidean vacuum on dS 
\cite{allen,chernikov-tagirov,tagirov,birrell-davies,laf-eucl}
and discuss the induced vacuum on $\rpdes$.

\subsection{Euclidean vacuum on de~Sitter}

The massive scalar field action on a general curved
spacetime is
\begin{equation}
S = - \casehalf \int \sqrt{-g} \, d^4x
\left[ g^{\mu\nu} \phi_{,\mu} \phi_{,\nu}
+ ( \fieldmass^2 + \xi R) \phi^2
\right]
\ \ ,
\label{scalar-action-gen}
\end{equation}
where $\fieldmass$ is the mass, $R$ is the Ricci scalar, and $\xi$ is
the curvature coupling constant. Specializing to~dS, we have $R = 12
H^2$.  We assume $\fieldmass^2 + 12\xi H^2>0$, and we define the
effective mass as $\fieldeffmass := \sqrt{\fieldmass^2 + 12\xi H^2}$.

The field equation reads 
\begin{equation}
\left(
\Box
- \fieldeffmass^2
\right)
\phi =0
\ \ ,
\label{field-equation}
\end{equation}
where $\Box$ stands for the scalar Laplacian on~dS\null. 
The (indefinite) inner product, evaluated on a hypersurface of
constant~$t$, is 
\begin{equation}
(\phi_1,\phi_2) :=
i H^{-3} \cosh^3(Ht) 
\int_{S^3} \sin^2\!\chi \, \sin\theta \, d\chi d\theta d\varphi
\>
\overline{\phi_1}
\,
\tensor{\partial}_t
\phi_2 \,
\ \ . 
\label{inner-product}
\end{equation}

The spatial dependence of the field equation (\ref{field-equation})
can be separated by the hyperspherical harmonics~$Q_{nlm}$, which are
eigenfunctions of the Laplacian on the unit three-sphere with the
eigenvalue $-(n^2-1)$: here $n = 1,2,\ldots\,$, and the degeneracy
described by the indices $l$ and $m$ is~$n^2$. For more detail about
the harmonics, see for example \cite{lif-khal-harm}. The
remaining, time-dependent equation can then be solved in terms of
associated Legendre functions. For the normalized positive frequency
mode functions, we choose
\begin{eqnarray}
\phi_{nlm} := 
&&
e^{-i\nu\pi/2}
\sqrt{
\frac{\pi H^2 \Gamma(n + \casehalf - \nu)}
{4\Gamma(n + \casehalf + \nu)}
}
\, 
Q_{nlm}
\times
\nonumber
\\
&&
\times
\sin^{3\over2}(\eta) 
\left[
{\rm P}^\nu_{n-{1\over2}} \left( -\cos\eta \right) 
- (2i/\pi) 
{\rm Q}^\nu_{n-{1\over2}} \left( -\cos\eta \right) 
\right]
\ \ ,
\label{des-CTmodes}
\end{eqnarray}
where ${\rm P}^\nu_{n-{1\over2}}$ and ${\rm Q}^\nu_{n-{1\over2}}$ are
the associated Legendre functions on the cut
\cite{abra-stegun,Grad-Rhyz,MOS} and $\nu$ is one of the solutions of
\begin{equation}
\nu^2 = \case{9}{4} - \fieldeffmass^2 H^{-2}
\ \ .
\label{nu-def}
\end{equation}
Which of the two solutions of (\ref{nu-def}) 
is chosen for $\nu$ is immaterial, 
as the two choices give equivalent mode functions.
The resulting vacuum, which we denote by~$\evacdes$, is known as the
Euclidean vacuum or the Chernikov-Tagirov vacuum
\cite{chernikov-tagirov,tagirov,birrell-davies}.

The vacuum $\evacdes$ is uniquely characterized by the properties that
its two-point function $G^+_{\rm dS}(x,x')$ is invariant under the
connected component of the isometry group of~dS, and that the only 
singularity of $G^+_{\rm dS} (x,x')$ occurs when
$x'$ is on the light cone of~$x$ \cite{allen}. 
Explicitly, we have 
\begin{equation}
G^+_{\rm dS} (x,x')
= A H^2 
{\tilde F}
\left(
\casehalf \! \left[ 1 + {\cal Z}_\epsilon (x,x') \right]
\right)
\ \ ,
\label{des-wightmanplus}
\end{equation}
where ${\tilde F}$ is the hypergeometric function 
\cite{abra-stegun} 
\begin{equation}
{\tilde{F}}(z)
:=
\hypergtwoone \left(\case{3}{2} + \nu, \case{3}{2} - \nu; 2;
z
\right)
\ \ ,
\label{Ftilde-def}
\end{equation}
and the numerical factor $A$ is given by 
\begin{equation} 
A :=  \frac{\fieldeffmass^2 H^{-2} - 2}{16\pi \cos \pi\nu}
\label{A-def}
\end{equation}
for $\fieldeffmass^2H^{-2} \ne 2$, 
and in the special case $\fieldeffmass^2H^{-2} = 2$
by the limiting value of~(\ref{A-def}), $A = 1/(16\pi^2)$. 
Here ${\cal Z}_\epsilon(x,y)$ is equal to ${\cal Z}(x,y)$
(\ref{calZed-def}), but understood near ${\cal Z}(x,y)=1$ in a sense
that gives $G^+_{\rm dS} (x,x')$ the correct singularity structure on
the light cone \cite{birrell-davies}: we can represent ${\cal
  Z}_\epsilon$ for example by
\begin{equation}
{\cal Z}_\epsilon (x,y) := 
{\cal Z}(x,y) 
- i \epsilon \left[ U(x) - U(y) \right]
- \epsilon^2 {\left[ U(x) - U(y) \right]}^2
\end{equation}
where $\epsilon\to0_+$. 

The renormalized stress-energy tensor in
$\evacdes$ is by construction invariant under the isometries
of~dS\null, and hence proportional to the metric tensor. 
In particular, the energy density measured by an inertial 
observer is constant along the observer trajectory and the same for
every observer. 
The explicit expression for the stress-energy tensor 
can be found for example in \cite{bd-des-T}. 

dS~can be regarded as a Lorentzian section of a complex spacetime
that admits the round four-sphere as a Riemannian section. The Feynman 
propagator in $\evacdes$ then analytically continues to the unique Green
function on the Riemannian section. This property is the origin of 
the name ``Euclidean vacuum'' for $\evacdes$.

\subsection{Euclidean vacuum on $\rpdes$}

The above quantization on dS adapts to $\rpdes$ with the obvious
modifications. In the inner product~(\ref{inner-product}), the spatial
integration is now over~$\RPthree$. The spatial dependence of the
field equation (\ref{field-equation}) is separated by the harmonics on
the unit~$\RPthree$: these harmonics are constructed by taking from
the set $\left\{Q_{nlm}\right\}$ those that are invariant under the
antipodal map, projecting to~$\RPthree$, and multiplying by $\sqrt{2}$
to achieve the correct normalization.\footnote{For more detail, see
  \cite{schleich-witt-rpthreedes}.} In the normalized positive
frequency mode functions, we choose the time dependence as
in~(\ref{des-CTmodes}). We denote the resulting vacuum
by~$\evacrpdes$.

As $\evacrpdes$ is induced by $\evacdes$ under the projection ${\rm
  dS}\to \rpdes$, the two-point functions in $\evacrpdes$ are obtained
from those in $\evacdes$ by the method of images. For example, for the
positive frequency Wightman function $G^+_{\rpdes}(x,x')$
in~$\evacrpdes$, we have 
\begin{equation}
G^+_{\rpdes}(x,x') =
G^+_{\rm dS}
(x,x')
+
G^+_{\rm dS}
\biglb(x,J(x') \bigrb)
\ \ . 
\label{Gplus-image}
\end{equation}
where $x$ and $x'$ on the two sides of the equation are understood as
points in dS or $\rpdes$ in the obvious way. It follows that all the
two-point functions in $\evacrpdes$ are invariant under the connected
component of the isometry group of $\rpdes$. 

$\rpdes$ can be regarded as a Lorentzian section of a complex
spacetime using the formalism of
(anti)holomorphic involutions \cite{gibb-holo,chamb-gibb}, and 
its Riemannian section can then be defined as a certain
$\BbbZ_2$ quotient of the round four-sphere. By method-of-images 
techniques similar to those used in \cite{louko-marolf-geon},
one sees that the Feynman propagator in $\evacrpdes$ analytically
continues to the unique Green function on the Riemannian section.  We
therefore refer to $\evacrpdes$ as the Euclidean vacuum on $\rpdes$.

{}Using~(\ref{Gplus-image}), it is straightforward to compute the
stress-energy tensor in $\evacrpdes$ by the point-splitting method
\cite{birrell-davies}. The contribution from the first term on the
right-hand side of (\ref{Gplus-image}) is identical to the
stress-energy tensor in~dS\null. The remaining contribution, 
arising from the second term on the right-hand side
of~(\ref{Gplus-image}), is finite without additional renormalization,
and it is clearly invariant under the isometries of $\rpdes$. 
Denoting this contribution by ${\Delta T}_{\mu\nu}$, 
we find that
its nonvanishing 
mixed components in the 
coordinates $(t,\chi,\theta,\varphi)$ are 
\begin{mathletters}
\label{DeltaT}
\begin{eqnarray}
{\Delta T}^t{}_t 
&=& 
A H^4 \left[
3\xi {\tilde{F}}(z) + \case{3}{2}(1-4\xi) z {\tilde{F}}'(z) 
\right]
\ \ ,
\\
{\Delta T}^i{}_j 
&=& 
A H^4
\left\{
\left[ 3\xi + (4\xi-1)\fieldeffmass^2 H^{-2} \right]
{\tilde{F}}(z) 
+ \left[\case{1}{2}(16\xi-3)z + 1 - 6\xi \right] 
{\tilde{F}}'(z) 
\right\}
\delta^i{}_j
\ \ ,
\end{eqnarray}
\end{mathletters}
where the Latin indices stand for the spatial coordinates 
$(\chi,\theta,\varphi)$, $z := -\sinh^2(Ht)$, and 
${\tilde{F}}'(z) := d{\tilde{F}}(z)/dz$. 

It is clear from (\ref{DeltaT}) that the energy density measured by an 
inertial observer is not constant along the observer
trajectory. However, it follows from the expansions of
hypergeometric functions \cite{abra-stegun} that all the components of 
${\Delta T}^\mu{}_\nu$ fall off exponentially in
$t$ at large~$|t|$, the details of the falloff depending on the
parameters. Therefore, in the distant past and future of each observer 
trajectory, the stress-energy tensor in $\evacrpdes$ is
exponentially asymptotic to the stress-energy tensor in $\evacdes$. 

As a special case, consider the massless conformally-coupled field, 
for which $\xi =
\case{1}{6}$ and $\mu=0$. Then $\nu=\case{1}{2}$, 
$A = 1/(16\pi^2)$,
and ${\tilde{F}}(z) = 1/(1-z)$. In the coordinates 
$(t,\chi,\theta,\varphi)$, we obtain 
\begin{equation}
{\Delta T}^\mu{}_\nu
= 
\frac{H^4}{32\pi^2 \cosh^4(Ht)}
\, 
{\rm diag} \, 
\left( 
1, -\case{1}{3}, -\case{1}{3}, -\case{1}{3} 
\right)
\ \ .
\label{deltaT-conformal}
\end{equation}
In the Appendix we independently verify the result
(\ref{deltaT-conformal}) by techniques that take
advantage of the conformal relation between $\rpdes$ and the
$\RPthree$ version of the Einstein static universe.

\section{Bogoliubov transformation on $\rpdes$}
\label{sec:bogo-transf}

In this section we use a Bogoliubov-transformation technique to
examine the experiences of an inertial observer in $\rpdes$, under the
assumption that the world line of the observer is normal to the
distinguished spacelike foliation. In subsection \ref{subsec:boulvac}
we review the quantization in the spacetime region covered by the
static coordinates, centered around the world line of the observer,
and we recall the construction of the Boulware-like
vacuum~$\boulvacdes$ in this region. In subsection
\ref{subsec:boul-eucl-bogo} we express $\evacrpdes$ in terms of
the excited states built on $\boulvacdes$ and interpret the result in
terms of particles seen by the observer.

\subsection{Quantization in the static coordinates}
\label{subsec:boulvac}

In this subsection we review the quantization of a real scalar field
$\phi$ in the spacetime 
covered by the static
metric~(\ref{static-metric}). 
As explained in section~\ref{sec:spacetimes}, 
we can interpret this spacetime as the
quadrant $\desquadrant$ in~dS, or as the region $\rpquadrant$ in
$\rpdes$. 

The (indefinite) inner product, 
evaluated on a hypersurface of
constant~$\statictime$, reads
\begin{equation}
(\phi_1,\phi_2) :=
i \int_{S^2} \sin\theta \, d\theta d\varphi
\int_0^{\infty}
r^2 dr^*
\,
\overline{\phi_1}
\,
\tensor{\partial}_\statictime
\phi_2
\ \ ,
\label{ext-inner-product}
\end{equation}
where $r^*$ is the tortoise coordinate,
\begin{equation}
r^* := \frac{1}{2H}
\ln \left( \frac{1 + Hr}{1 - Hr} \right)
\ \ ,
\end{equation}
having the range $0\le r^* <\infty$. 
Separating the field equation (\ref{field-equation}) by the ansatz 
\begin{equation}
\phi =
{(4\pi\omega)}^{-1/2}
r^{-1} R_{\omega l}(r) e^{-i\omega\statictime }
Y_{lm}(\theta,\varphi)
\ \ ,
\label{separ-ansatz}
\end{equation}
where $Y_{lm}$ are the spherical
harmonics\footnote{We use the
Condon-Shortley phase convention (see for example \cite{arfken}), 
in which
$Y_{l(-m)}(\theta,\varphi) =
{(-1)}^m \overline{Y_{lm}(\theta,\varphi)}$
and
$Y_{lm}(\pi-\theta,\varphi+\pi) =
{(-1)}^l Y_{lm}(\theta,\varphi)$.}, 
the equation for the radial function $R_{\omega l}(r)$ becomes 
\begin{equation}
0 =
\left[
\frac{d^2}{d {r^*}^2}
+ \omega^2
- \left( 1 - H^2r^2 \right)
\left( \fieldeffmass^2 - 2H^2 + \frac{l(l+1)}{r^2}
\right)
\right]
R_{\omega l}
\ \ .
\label{radial-eq}
\end{equation}
The one-dimensional differential operator in (\ref{radial-eq}) is
essentially self-adjoint with respect to the Schr\"odinger-type inner
product $\int_0^\infty dr^* \overline{R_1} R_2$ for $l>0$, and for
$l=0$ we choose for this operator the self-adjoint extension whose
(generalized) eigenfunctions vanish at $r^*=0$. The spatial parts of
the wave functions (\ref{separ-ansatz}) are then the (generalized)
eigenfunctions 
of the essentially self-adjoint spatial part of the wave operator in
the field equation~(\ref{field-equation}) \cite{gal-pas-laplacian},
which in particular means
that $R_{\omega l}$ are at small $r^*$ proportional to $(r^*)^{l+1}$.
It follows by standard techniques\footnote{When $l=0$ and
  $\fieldeffmass^2 < 2H^2$, equation (\ref{radial-eq}) is analyzed for
  example in \cite{lan-lif-pot}. In other cases the analysis is
  standard by the nonnegativity of the potential term
  in~(\ref{radial-eq}).}
that for each~$l$, the spectrum of
$\omega^2$ is continuous and spans the positive real axis.

We choose the positive frequency mode functions to have $\omega>0$,
and we denote the resulting vacuum by $\boulvacdes$. As these mode
functions are positive frequency with respect to the timelike Killing
vector~$\partial_\statictime$, which generates the inertial motion
along the geodesic at $r=0$, an observer moving along this geodesic
experiences $\boulvacdes$ as her physical no-particle state:
$\boulvacdes$ is analogous to the Boulware vacuum on the exterior
Schwarzschild, and to the Rindler vacuum in a Rindler wedge on
Minkowski space.  For a complete orthonormal set of positive frequency
modes, we choose
\begin{equation}
u_{\omega lm}
:=
e^{i(l+|m|)\pi/2}
{(4\pi\omega)}^{-1/2}
r^{-1} R_{\omega l} e^{-i\omega \statictime}
Y_{lm}
\ \ ,
\label{boulware-modes}
\end{equation}
where the functions $R_{\omega l}$ are real-valued and normalized so
that their asymptotic form at large $r^*$ is 
\begin{equation}
R_{\omega l} \sim 
2 \cos(\omega r^* + \delta_{\omega l})
\ \ ,
\ \ 
r^* \to \infty
\ \ ,
\end{equation}
where $\delta_{\omega l}$ is a real-valued phase shift. 
The orthonormality relation reads
\begin{equation}
(u_{\omega lm} ,u_{\omega' l' m'} )
=
\delta_{l l'}
\delta_{m m'}
\delta(\omega-\omega')
\ \ ,
\end{equation}
with the complex conjugates satisfying a similar relation with a
minus sign, and the mixed inner products vanishing.

We expand the quantized field as
\begin{equation}
\phi =
\sum_{l m }
\int_0^\infty d\omega
\left[
b_{\omega lm} u_{\omega lm}
+
b_{\omega lm}^\dagger
\overline{u_{\omega lm}}
\right]
\ \ ,
\label{phi-static-expansion}
\end{equation}
where $b_{\omega lm}$ and
$b_{\omega lm}^\dagger$
are the annihilation and creation operators associated with the
mode $u_{\omega lm}$. The vacuum 
$\boulvacdes$ satisfies by definition 
\begin{equation}
b_{\omega lm} \boulvacdes = 0
\ \ .
\end{equation}

\subsection{Bogoliubov transformation}
\label{subsec:boul-eucl-bogo}

We now consider the above quantization in the static coordinates as
having been performed in the region $\rpquadrant$ of $\rpdes$. 
We wish to write the vacuum induced on $\rpquadrant$ by $\evacrpdes$ in
terms of $\boulvacdes$ and the excitations created by 
$\left\{b_{\omega lm}^\dagger\right\}$. 
Much of our analysis builds on the transformations 
developed for dS in \cite{lapedes}. 

Rather than computing directly the Bogoliubov transformation between
the sets $\left\{\phi_{nlm}\right\}$ (\ref{des-CTmodes}) and
$\left\{u_{\omega lm}\right\}$~(\ref{boulware-modes}), we take
advantage of the observation that the modes $\phi_{nlm}$ are analytic
functions in $\eta$ in the lower half of the strip
$0<{\rm{Re}}(\eta)<\pi$ in the complex $\eta$ plane, and that they are
bounded as ${\rm{Im}}(\eta)\to-\infty$ in this strip \cite{MOSexp}.
Following Unruh \cite{unruh-magnum}, we can therefore find a set of
modes that share the vacuum $\evacrpdes$ by forming from
$\left\{u_{\omega lm}\right\}$ and their complex conjugates 
linear combinations that are analytically continued across the horizons
with ${\rm{Im}}(\eta)<0$, and globally well defined 
on $\rpdes$.  We call these modes $W$-modes.

The construction of the $W$-modes follows closely that 
in the Rindler-type spacetime in 
\cite{louko-marolf-geon}. We coordinatize $\rpdes$ by $(\eta, \chi,
\theta, \varphi)$ in the sense explained in
section~\ref{sec:spacetimes}. The region of $\rpdes$ covered by the
static coordinates $(\statictime, r, \theta, \varphi)$ is then
$\chi-\casehalf\pi < |\eta-\casehalf\pi|$, and the embedding is given
by by~(\ref{transf-charts}). Near the horizon in the static region,
$\chi-\casehalf\pi \to |\eta-\casehalf\pi|$, $u_{\omega lm}$ is
asymptotically 
proportional to
\begin{equation}
\left\{
e^{i\delta_{\omega l}}
{\left[
\tan\left(\frac{\eta-\chi}{2}\right) 
\right]}^{-i\omega/H}
+ 
e^{-i\delta_{\omega l}}
{\left[
\tan\left(\frac{\eta+\chi}{2}\right) 
\right]}^{-i\omega/H}
\right\}
Y_{lm}
\ \ .
\label{bmode-nearhor}
\end{equation}
Continuing the asymptotic expression (\ref{bmode-nearhor}) past the
horizon $\eta=\chi$ into the past region of $\rpdes$, in the lower
half-plane in~$\eta$, we obtain
\begin{equation}
\left\{
e^{i\delta_{\omega l}}
e^{-\pi\omega/H}
{\left[
\tan\left(\frac{\chi-\eta}{2}\right) 
\right]}^{-i\omega/H}
+ 
e^{-i\delta_{\omega l}}
{\left[
\tan\left(\frac{\chi+\eta}{2}\right) 
\right]}^{-i\omega/H}
\right\}
Y_{lm}
\ \ .
\label{bmode-nearhor-past}
\end{equation}
In order to have the asymptotic form of a mode that is globally
well defined in the past region, one needs to add to
(\ref{bmode-nearhor-past}) its image under~(\ref{J-def}), which is 
\begin{equation}
{(-1)}^l 
\left\{
e^{i\delta_{\omega l}}
e^{-\pi\omega/H}
{\left[
\tan\left(\frac{\chi+\eta}{2}\right) 
\right]}^{i\omega/H}
+ 
e^{-i\delta_{\omega l}}
{\left[
\tan\left(\frac{\chi-\eta}{2}\right) 
\right]}^{+i\omega/H}
\right\}
Y_{lm}
\ \ .
\label{bmode-nearhor-past-image}
\end{equation}
Continuing the sum of (\ref{bmode-nearhor-past}) and
(\ref{bmode-nearhor-past-image}) back to the static region, matching
the asymptotic form to a linear combination from the set
$\left\{u_{\omega lm}\right\}$, and 
normalizing, we recover the modes 
\begin{equation}
W_{\omega l m}
:=
{1 \over \sqrt{2 \sinh(\pi\omega/H)}}
\left(
e^{\pi\omega/2 H } \,
u_{\omega lm}
+
e^{- \pi\omega/2 H} \,
\overline{u_{\omega l(-m)} }
\right)
\ \ .
\label{W-modes-rpdes}
\end{equation}
A continuation to and from the future region of $\rpdes$ instead of
the past region is similar and leads again to~(\ref{W-modes-rpdes}).
The set $\left\{W_{\omega l m}\right\}$ provides the desired complete
orthonormal set of $W$-modes.

The quantized field can be expanded in terms of the $W$-modes as
\begin{equation}
\phi =
\sum_{l m }
\int_0^\infty d\omega
\left(
d_{\omega l m} W_{\omega l m}
+
d^\dagger_{\omega l m} \overline{W_{\omega l m}}
\right)
\ \ ,
\label{phi-W-expansion}
\end{equation}
where $d_{\omega l m}$ and $d^\dagger_{\omega l m}$
are respectively the annihilation and creation operators
associated with the mode $W_{\omega l m}$.
The vacuum of the $W$-modes is by
construction~$\evacrpdes$,
\begin{equation}
d_{\omega l m} \evacrpdes =0
\ \ .
\end{equation}
Comparing the expansions (\ref{phi-static-expansion})
and~(\ref{phi-W-expansion}), and using~(\ref{W-modes-rpdes}), we see
that the Bogoliubov transformation between the operators reads 
\begin{equation}
b_{\omega l m} =
{1 \over \sqrt{2 \sinh(\pi\omega/H)}}
\left(
e^{\pi\omega/2H} \,
d_{\omega l m} +
e^{-\pi\omega/2H} \,
d^\dagger_{\omega l (-m)}
\right)
\ \ .
\end{equation}
Suppressing $\omega$ and~$l$, and proceeding
as in \cite{louko-marolf-geon}, we obtain 
\begin{eqnarray}
\evacrpdes &=&
{1\over \sqrt{\cosh(r_\omega)}}
\left(
\sum_{q=0}^\infty
{(2q-1)!! \, \exp(-\pi \omega q/H)
\over \sqrt{(2q)!}}
|2q\rangle_0
\right)
\nonumber
\\
&&
\times
\prod_{m>0}^\infty
\left(
{1 \over \cosh(r_\omega)}
\sum_{q=0}^\infty
\exp(-\pi \omega q/H)
|q\rangle_m |q\rangle_{(-m)}
\right)
\ \ ,
\label{evacrpdes-expanded}
\end{eqnarray}
where 
\begin{equation}
\tanh(r_\omega) := \exp(-\pi\omega/H)
\ \ ,
\end{equation}
and $|q\rangle_m$ denotes the normalized state with $q$
excitations in the static mode labeled by $m$ (and the suppressed
indices $\omega$ and~$l$),
\begin{equation}
|q\rangle_n
:=
(q!)^{-1/2}
\left( b_m^\dagger \right)^q
\boulvacdes
\ \ .
\end{equation}
The notation in (\ref{evacrpdes-expanded}) is adapted to the tensor
product structure of the Hilbert space over the modes: the state
$|q\rangle_m |q\rangle_{(-m)}$ contains $q$ excitations both in the
mode $m$ and in the mode~$-m$. The vacuum $\evacrpdes$ therefore
contains excitations with $m\ne0$ in pairs whose members
only differ in the sign of~$m$.

For generic operators with support in the static region, or even with
support only on the inertial trajectory at $r=0$, the expectation
values in $\evacrpdes$ are clearly not thermal. However, suppose that
$\hat{A}$ is an operator with support in the static region, such that
$\hat{A}$ does not couple to the modes $u_{\omega l m}$ with $m=0$,
and for each triplet $(\omega l m)$ with $m\ne0$, $\hat{A}$ only
couples to one of the modes $u_{\omega l m}$ and $u_{\omega l(-m)}$.
It is easily seen from~(\ref{evacrpdes-expanded}), as in the
Rindler-analysis in \cite{louko-marolf-geon}, that the
expectation values of $\hat{A}$ are thermal in the temperature $T =
H/(2\pi)$.

The mode functions $u_{\omega l m}$ are unlocalized in~$\statictime$.
However, it is straightforward to adapt the above analysis to wave
packets partially localized in both $\statictime$ and $\omega$, as in the
Rindler case discussed in \cite{louko-marolf-geon}. In the
static region, one finds thermal expectation values in the temperature
$T = H/(2\pi)$ for any operator 
whose support is localized at
asymptotically early or late values of~$\statictime$.

\section{Inertial particle detector in $\evacrpdes$}
\label{sec:detector}

We now turn to the experiences of an inertial monopole particle
detector \cite{birrell-davies,unruh-magnum,dewitt-CC,takagi-magnum} in
$\rpdes$ in the vacuum~$\evacrpdes$.  As $\evacrpdes$ is invariant
under the 
isometries of $\rpdes$, we can without loss of
generality consider the detector 
trajectory whose one preimage on the
hyperboloid (\ref{des-hyperboloid}) is
\begin{mathletters}
\label{detector-trajectory}
\begin{eqnarray}
&&U = 
H^{-1} 
\sinh(H\tau) \cosh \gamma 
\ \ ,
\\
&&V = 
H^{-1} 
\cosh(H\tau) 
\ \ ,
\\
&&Z = 
H^{-1} 
\sinh(H\tau) \sinh \gamma 
\ \ ,
\\
&&X = 0 
\ \ ,
\\
&&Y = 0 
\ \ , 
\end{eqnarray}
\end{mathletters}
where $\gamma$ is a nonnegative parameter and 
$\tau$ is the proper time along the trajectory. 
Geometrically, 
$\gamma$ is the hyperbolic angle between the trajectory
at $\tau=0$ and the 
normal to the spacelike hypersurface belonging to the 
distinguished foliation of $\rpdes$. 
For $\gamma=0$, the trajectory is orthogonal to the 
the distinguished foliation at all~$\tau$. For $\gamma>0$, the
trajectory is nowhere orthogonal to this foliation, but it becomes
asymptotically orthogonal as $|\tau|\to\infty$. 
We shall consider on a par both this trajectory in 
$\rpdes$ and the well-known case of the trajectory
(\ref{detector-trajectory}) in~dS\null.

In first order perturbation theory, the probability for the detector
becoming excited is
\cite{birrell-davies,unruh-magnum,dewitt-CC,takagi-magnum}
\begin{equation}
c^2 \sum_{E>0}
{\left|
\vphantom{|^A_A}
\langle\!\langle E | \bbox{m}(0) | 0 \rangle\!\rangle \right|}^2
{\cal{F}}(E)
\ \ ,
\label{exci-prob}
\end{equation}
where $c$ is the coupling constant,
$\bbox{m}(\tau)$ is the detector's
monopole moment operator,
$| 0 \rangle\!\rangle$ is the ground state of the detector,
the sum is over all the excited states
$| E \rangle\!\rangle$ of the detector, and
the detector response function ${\cal{F}}(E)$ is given by 
\begin{equation}
{\cal{F}} (E)
:=
\int d\tau \int d\tau' \,
e^{-iE(\tau-\tau')}
G^+ \biglb( x(\tau), x(\tau') \bigrb)
\ \ .
\label{resp-function}
\end{equation}
For the trajectory (\ref{detector-trajectory}) in~dS, equation 
(\ref{des-wightmanplus}) yields 
\begin{equation}
G^+_{\rm dS} \biglb( x(\tau), x(\tau') \bigrb)
=
A H^2 
{\tilde F} \left(
\cosh^2 [H(\tau-\tau')/2 -i\epsilon]
\right)
\ \ .
\label{Gplustau-des}
\end{equation}
For the trajectory in $\rpdes$, equations (\ref{des-wightmanplus}) and
(\ref{Gplus-image}) give  
\begin{equation}
G^+_{\rpdes} \biglb( x(\tau), x(\tau') \bigrb)
=G^+_{\rm dS}
\biglb( x(\tau),
x(\tau') \bigrb)
+
\Delta G^+(\tau,\tau')
\ \ ,
\label{Gplustau-twoterms}
\end{equation}
where 
\begin{equation}
\Delta G^+(\tau,\tau')
= 
A H^2 
{\tilde F} 
\left(
\casehalf \! \left[ 1 + {\tilde{\cal Z}}_\epsilon(\tau,\tau')
\right]
\right)
\label{DeltaGplus-tau}
\end{equation}
with 
\begin{equation}
{\tilde{\cal Z}}_\epsilon(\tau,\tau')
:= - \cosh[H(\tau+\tau')]
- 2 \sinh^2 \!  \gamma \, \sinh(H\tau) \sinh(H\tau')  
- i \epsilon(\tau-\tau')
\ \ .
\label{tildeZtaus}
\end{equation}
For $\gamma=0$, the imaginary part in 
(\ref{tildeZtaus}) can be dropped, as the argument of ${\tilde F}$ in 
(\ref{DeltaGplus-tau}) is then always negative: the geometrical reason 
is that in this case the trajectory 
(\ref{detector-trajectory}) in dS and its image under 
$J$ have a spacelike separation, which guarantees that
$\Delta G^+(\tau,\tau')$ is necessarily nonsingular. 
For $\gamma>0$, on the other hand, the imaginary part in
(\ref{tildeZtaus}) is needed to specify the singularity structure in
$\Delta G^+(\tau,\tau')$ when the argument of ${\tilde F}$ in
(\ref{DeltaGplus-tau}) takes the value~$1$.

Consider now the familiar case of the 
detector (\ref{detector-trajectory}) in~dS\null. 
$G^+_{\rm dS} \biglb( x(\tau), x(\tau') \bigrb)$ 
(\ref{Gplustau-des}) is independent
of~$\gamma$, and it depends on $\tau$ and $\tau'$ only through the
difference $\tau-\tau'$, as the case must be by the invariance of
$\evacdes$ under the connected isometries
of~dS\null. 
If the detector is adiabatically turned on 
in the asymptotic past and off
in the asymptotic future, the total
response function ${\cal{F}}_{\rm dS}(E)$
is infinite, which reflects the fact that the
excitation rate is constant and nonvanishing along the trajectory: the
excitation rate in unit proper time is recovered by leaving out one of
the integrals in~(\ref{resp-function}). For 
$\fieldeffmass^2H^{-2}=2$, one recovers for the excitation rate 
the Planckian result at the
temperature $T = H/(2\pi)$ \cite{birrell-davies},
\begin{equation}
\frac{{\cal{F}}_{\rm dS}(E)}{\hbox{(unit proper time)}}
=
{E \over 2\pi
\left( e^{2\pi E/H} -1 \right)
}
\ \ .
\label{dS-Ftildef}
\end{equation}

Consider then the detector in $\rpdes$. As $\Delta G^+(\tau,\tau')$
(\ref{DeltaGplus-tau}) depends on $\tau$ and $\tau'$ not only through
the difference $\tau-\tau'$ but also through the individual values,
the excitation probability per unit proper time is not a constant
along the trajectory, and this probability depends also on~$\gamma$.
The detector therefore senses the distinction between the vacua
$\evacdes$ and $\evacrpdes$, and it also senses its velocity with
respect to the distinguished foliation of $\rpdes$.  However, if
$\tau$ and $\tau'$ are both large and positive, or if they are both
large and negative, ${\tilde{\cal Z}}_\epsilon(\tau,\tau')$
(\ref{tildeZtaus}) is large and negative, and $\Delta G^+(\tau,\tau')$
tends to zero as $|\tau+\tau'|\to\infty$ \cite{abra-stegun}. In the
the asymptotic future, or in the asymptotic past, the detector
therefore responds as in~$\evacdes$. For $\gamma=0$, this is the
result one would have expected from the Bogoliubov transformation of
section~\ref{sec:bogo-transf}.

\section{Summary and discussion}
\label{sec:discussion}

We have shown that the Euclidean vacua 
of a free scalar field on the spacetimes dS and $\rpdes$ are
distinguishable to an inertial observer who couples to the field
through a monopole detector, or to an observer who can measure the
field stress-energy tensor. In the special case of an inertial
observer whose world line on $\rpdes$ is orthogonal to the
distinguished foliation, we arrived at a similar conclusion by
constructing the Bogoliubov transformation between the modes that
define the Euclidean vacuum and the modes that are of positive
frequency with respect to the observer's natural time
coordinate. However, we also saw that the differences between
dS and $\rpdes$ become exponentially small in the
distant past or future on an inertial observer world line, and in
these limits the observer thus sees the Euclidean vacuum on $\rpdes$
as a thermal bath in the usual de~Sitter Hawking temperature.  This
result conforms to the central tenet of inflationary cosmology,
namely, that the physics in an exponentially expanding spacetime
should become
indistinguishable from physics in de~Sitter
space exponentially fast: what falls off exponentially 
in our case are the effects of the unconventional spatial topology
on the quantum field. 

While our particle detector analysis accommodated an arbitrary
inertial observer in $\rpdes$, we only performed the Bogoliubov
transformation for an inertial observer whose world line is orthogonal
to the distinguished foliation. As any inertial world line in $\rpdes$
has a neighborhood covered by the static metric~(\ref{static-metric}),
such that the world line is at $r=0$, the case of a nonorthogonal
trajectory would also be in principle amenable to a Bogoliubov
transformation analysis. 
One would expect the correlations in the counterpart of
(\ref{evacrpdes-expanded}) to be more complicated for a nonorthogonal
trajectory, but one would expect a wave
packet analysis to show thermality in the limit of early and
late times also in this case. Finding the Bogoliubov
transformation explicitly for a nonorthogonal trajectory 
is however more difficult, and we
shall not pursue this question further here. 

Yet another way to investigate the experiences of an inertial observer 
in $\rpdes$ is through the complex analytic properties of the
Feynman propagator in $\evacrpdes$. 
As mentioned in section~\ref{sec:quantization}, $\rpdes$ can regarded
as a Lorentzian section of a complex spacetime whose Riemannian
section is a certain $\BbbZ_2$ quotient of the round four-sphere, 
and the Feynman propagator in $\evacrpdes$ continues to the unique
Green function on this Riemannian section. A set of coordinates
covering the Riemannian section can be obtained from the
static coordinates 
$(\statictime,r,\theta,\varphi)$ on $\rpquadrant$ by setting 
$\statictime = -i\estatictime$, provided the coordinates 
$(\estatictime,r,\theta,\varphi)$ are 
identified as 
\begin{equation}
(\estatictime,r,\theta,\varphi) 
\sim
(\estatictime + 2\pi/H,r,\theta,\varphi) 
\sim
(\pi/H  - \estatictime,r,\pi-\theta,\varphi+\pi)
\ \ .
\label{riem-idents}
\end{equation}
The first identification in (\ref{riem-idents}) is just as for~dS, and
this identification implies
for the Feynman propagator in $\evacdes$ complex
analytic properties that correspond to thermality in the 
de~Sitter Hawking temperature $H/(2\pi)$
\cite{GH-cosmo}. The second identification in (\ref{riem-idents}) 
is specific to $\rpdes$. One can argue that the complex analytic
properties of the Feynman propagator in $\evacrpdes$ are consistent
with thermality in the limit of asymptotically early and late proper
times: the reasoning is similar to that given for the $\RPthree$ geon
in \cite{louko-marolf-geon}, and we shall not spell out the
detail here. 

The action of the Riemannian section of $\rpdes$ is half of the action
of the Riemannian section of~dS\null. If one uses these actions
in a semiclassical estimate to a quantum gravitational
partition function, and if one associates to $\rpdes$ the de~Sitter
Hawking temperature~$H/(2\pi)$, one finds for the entropy of $\rpdes$
the result $2\pi H^{-2}$, which is half of the value obtained for dS
\cite{GH1}.  An analogous observation was made in 
\cite{louko-marolf-geon} for the entropy of the $\RPthree$ geon.
Although the entropy associated with
cosmological horizons may be physically 
less clear than the entropy associated with
black hole horizons, it should 
prove interesting to understand whether
this naive instanton-method evaluation of the entropy of $\rpdes$
could be physically justified, and in particular whether the factor
half relative to dS might also arise in any state-counting approach to
the entropy.

\acknowledgments
We would like to thank 
John Friedman, 
Ted Jacobson,
Don Marolf, 
Max Niedermaier,
and Don Witt
for discussions and correspondence.

\appendix
\section*{Stress-energy tensor via a conformal transformation} 

In this appendix we verify the result (\ref{deltaT-conformal}) for the
stress-energy tensor of a massless conformally coupled field by the
conformal technique of Parker \cite{Parker} and zeta-function
regularization.

To begin, recall \cite{Parker} that any 
covariantly conserved symmetric tensor $K^{\mu\nu}$ 
in a spacetime with a conformal 
Killing vector\footnote{A conformal Killing vector in four 
spacetime dimensions satisfies 
$\nabla_\mu\xi_\nu+\nabla_\nu\xi_\mu = \frac 12 \nabla_\rho
\xi^\rho g_{\mu\nu}$.}
$\xi^\nu$ satisfies the relation
\begin{equation}
\nabla_\mu(K^{\mu\nu}\xi_\nu)=\frac 14 \nabla_\nu
\xi^\nu K^\mu_\mu
\ \ .
\label{covcons}
\end{equation}
Integration of (\ref{covcons})
over a compact spacetime region $M$ with spacelike boundary 
$\partial M$ yields 
\begin{equation}
\int_{\partial M} d^3x \, h^\frac 12 K^{\mu\nu}\xi_\nu 
n_\mu = 
 \frac 14 \int_M  d^4x (-g)^\frac 12 
 \nabla_\nu\xi^\nu K^\mu_\mu
\ \ ,
 \label{gausslaw}
\end{equation}
where $n_\mu$ is the outward unit normal form on~$\partial M$.
If both the spacetime and $K^{\mu\nu}$ are spatially
homogeneous, we can choose $\partial M$ to consist of two 
homogeneous
spatial hypersurfaces, and the spatial integration in (\ref{gausslaw})
then factors out on both sides of the equation. One recovers a
relation that relates the projection of $K^{\mu\nu}$ orthogonal to the 
homogeneity hypersurfaces to~$K^\mu_\mu$. 

We apply the above 
to the spacetimes dS and $\rpdes$, 
for both of which the metric takes the form~(\ref{des-glob-metric}). 
For the conformal Killing vector~$\xi^\mu$, we choose 
$\partial_\eta  = H^{-1}\cosh (Ht) \partial_t$, for which 
$\nabla_\nu\xi^\nu = 4 \sinh(Ht)$. 
In the coordinates $(t,\chi,\theta,\varphi)$, we obtain 
\begin{equation}
\cosh^4(Ht_f)K^{00}(t_f) -
\cosh^4(Ht_i)K^{00}(t_i)= 
- H   \int_{t_i}^{t_f} dt \, \cosh^3(Ht)\sinh(Ht)
K^\mu_\mu(t)
\ \ .
\label{stressenergy}
\end{equation}

We wish to use the relation (\ref{stressenergy}) to determine the
difference of the renormalized stress-energy tensors in the vacua
$\evacdes$ and $\evacrpdes$. We denote these tensors respectively
by 
$\langle T_\s3^{\mu\nu}\rangle$ and 
$\langle T_\RPthree^{\mu\nu}\rangle$. 
Both tensors 
are covariantly conserved and invariant 
under the 
isometries of the respective spacetimes. 
Further, we can use the projection from 
dS to $\rpdes$ to map $\langle T_\s3^{\mu\nu}\rangle$
into a tensor on $\rpdes$. 
By the usual abuse of notation, we denote also this tensor on $\rpdes$ 
by $\langle T_\s3^{\mu\nu}\rangle$. 
Then $\Delta T^{\mu\nu}:=
\langle T_\RPthree^{\mu\nu}\rangle-\langle T_\s3^{\mu\nu}\rangle$ 
is a well-defined, 
covariantly-conserved tensor on $\rpdes$, 
and it fully characterizes the
differences in the stress-energy tensors in $\evacdes$ and
$\evacrpdes$. 
Equation (\ref{stressenergy}) hence holds with 
$K^{\mu\nu} = \Delta T^{\mu\nu}$. 

We now specialize to the conformally coupled massless field, 
$\mu=0$ and $\xi=\case{1}{6}$. 
As the divergences
in the trace of the renormalized stress energy tensor 
are purely local and anomalous, they
are determined entirely by the local geometry. These 
contributions are the same for $\rpdes$ and~dS; thus
$\Delta T^\mu_\mu= 0$. Equation (\ref{stressenergy}) 
with 
$K^{\mu\nu} = \Delta T^{\mu\nu}$ 
then implies 
\begin{equation}
\Delta T^{00}(t)=\frac C{\cosh^4(Ht)}
\ \ ,
\label{DeltaT00}
\end{equation}
where $C$ is a constant. Together with the tracelessness and
symmetries of $\Delta T^{\mu\nu}$, this implies 
\begin{equation}
{\Delta T}^\mu{}_\nu
= 
\frac{C}{\cosh^4(Ht)}
\, 
{\rm diag} \, 
\left( 
-1, \case{1}{3}, \case{1}{3}, \case{1}{3} 
\right)
\ \ .
\label{deltaT-conformal-C}
\end{equation}
Note that $\Delta T^{00}$ behaves as if it were classical
radiation. In particular, it 
redshifts exponentially to zero at large times~$t$. 

To evaluate the constant~$C$, and in particular to 
show that it is nonzero\footnote{This point is nontrivial: 
see Kennedy and Unwin \cite{KennedyandUnwin}
for an example where changing the boundary conditions on states does
not change the energy density.}, 
we employ a conformal transformation
technique. Observe \cite{bd-conformal}
that $\Delta T^{00}$
is entirely due to the nongeometrical
contribution from the conformal vacuum, {\em i.e\/}., 
that reflecting the boundary conditions on the state
set by the topology rather than that  
contributed from the anomalous trace.
One can therefore compute $\Delta T^{00}$ by first finding the
corresponding quantity, $\Delta \tilde T^{00}$, 
in suitable conformally 
related spacetimes and then
performing a conformal transformation: 
from equation (6.129) in \cite{bd-conformal}, this transformation reads 
\begin{equation}
\Delta T^\nu_\mu =\left(\frac {\tilde g}g\right)^\frac 12 \Delta 
\tilde T^\nu_\mu
\ \ ,
\label{cdensity}
\end{equation}
where $g$ and $\tilde g$ are the determinants of the 
conformally related metrics. 

Suitable spacetimes conformally related to dS and $\rpdes$ are
respectively the usual Einstein static
universe, with spatial topology~$S^3$, 
and the Einstein static universe with spatial topology
$\RPthree$. Their metrics are obtained by multiplying the metric
(\ref{des-conf-metric}), on 
respectively dS and $\rpdes$, by $\sin^2\!\eta$. 
For the ordinary Einstein static universe 
with curvature radius~$c$, 
the energy density is $\rho_\s3= 1/(480\pi^2 c^4)$ 
\cite{ford,dow-cr}, and a derivation of this result by zeta-function
regularization methods \cite{dow-ban} is given in
\cite{bd-estatic-energy}. For the Einstein static universe with
spatial topology $\RPthree$, we adapt the zeta-function 
calculation of 
\cite{bd-estatic-energy}, noting that among the hyperspherical
harmonics $\left\{Q_{nlm}\right\}$ on the round $S^3$, the ones that
project to the round $\RPthree$ are precisely those whose principal
quantum number $n$ is odd
\cite{schleich-witt-rpthreedes}. 
The regularized expression for the total energy on the spacelike
hypersurfaces of the 
Einstein static universe with
spatial topology $\RPthree$ and curvature radius $c$ 
reads thus 
\begin{equation}
E(s)_\RPthree = 
\casehalf 
\sum _{{n\ge1} \atop { n \ {\rm odd}}} n^2 
{(n/c)}^{-s}
=
\case{1}{2} c^s 
\left( 1 - 2^{2-s} \right) 
\zeta(s-2)
\ \ ,
\end{equation}
where $s$ is the regularization parameter and $\zeta$ is the Riemann
zeta-function. 
Taking $s=-1$  and dividing by the spatial volume 
$\pi^2c^3$ yields the 
energy density  
$\rho_\RPthree = -7/(240\pi^2c^4)$. 
Taking the difference between 
$\rho_\RPthree$ and~$\rho_\s3$, 
we obtain 
\begin{equation}
\Delta \tilde T^{00} = -\frac 1{32\pi^2 c^4}
\ \ .
\label{this}
\end{equation}
{}From (\ref{cdensity}) and 
(\ref{this}) we thus find 
\begin{equation}
 \Delta T^{00}(t) = -\frac {H^4}{32\pi^2\cosh^4(Ht)}
\ \ .
\label{theresult}
\end{equation}
which is (\ref{DeltaT00}) with 
$C = -H^4/(32\pi^2)$. With this value of~$C$, the expression
(\ref{deltaT-conformal-C}) agrees with the result
(\ref{deltaT-conformal}) obtained in the main 
text by point-splitting methods. 

If the field is not conformally coupled and massless, 
$\Delta T^\mu_\mu$ need not
vanish. It would be possible to obtain partial information 
about $\Delta T^{00}$, in particular about its falloff as
$|t|\to\infty$, by first computing $\Delta T^\mu_\mu$ via 
point-split methods and then applying~(\ref{stressenergy}). 
This calculation would, however, not be substantially simpler than the
full point-split evaluation of $\Delta T^{\mu\nu}$.

\newpage

\end{document}